  \documentclass{aa}

  \def\swift{{\it Swift/Gehrels~}} 
  \def\chandra{{\it Chandra~}} 
  \def\nustar{{\it NuSTAR~}} 
  \def\funits{$\rm erg\,cm^{-2}\,s^{-1}$~}

  \usepackage{natbib}
  \usepackage{graphicx}

  \usepackage{txfonts}

\begin{document}

     \title{The source number counts at high energies: SWIFT vs. NuSTAR}

     
      \titlerunning{The hard X-ray universe's number count distributions}
      \authorrunning{A. Akylas et al.}
     
     \author{A. Akylas
            \inst{1}
            \and
            I. Georgantopoulos
            \inst{1}
                        }

     \institute{Institute for Astronomy Astrophysics Space Applications and Remote Sensing (IAASARS), National Observatory of Athens, I. Metaxa \& V. Pavlou 1, Penteli, 15236, Greece\\
                \email{aakylas@noa.gr}
                      }

    \date{}

    \abstract{
      The hard X-ray sky at energies above 10 keV, has been extensively explored by the \swift  and the {\it NuSTAR} missions in the  14-195 keV and the 3-24 keV bands respectively. 
      The major population of the hard X-ray detected sources are Active Galactic Nuclei (AGN). A discrepancy has been reported
      between the number count distributions of the two missions in the sense that the extrapolation of the \swift number counts
       in the flux regime sampled by {\it NuSTAR} lies significantly below the {\it NuSTAR} counts. 
      We explore anew this issue by comparing the number count distributions derived from the \swift 105-month catalogue 
      with those from the serendipitous {\it NuSTAR} source catalogue. 
      We use a novel number count distribution estimator which makes use of the C-statistic on the un-binned data. 
      In the 14-195 keV band, the {\it Swift}/BAT counts follow a Euclidean slope with $\alpha=1.51\pm0.10$ (90\% confidence level). 
      The {\it NuSTAR} counts in the 8-24 keV band present  a steeper slope with  $\alpha = 1.71\pm 0.20$, because of an upturn at fluxes below $\sim2\times10^{-13}$ \funits.
      The same upturn is observed in the soft (3-8 keV) {\it NuSTAR} number counts, which in overall also  present a steep slope with $\alpha = 1.82\pm 0.15$. 
      Only the bright part of the {\it NuSTAR} 3-8 keV counts agrees with the \chandra number counts in the 2-10 keV band while 
      the fainter part (below $\sim 7\times10^{-13}$ \funits ) of the soft \nustar counts is in marked disagreement with the \chandra counts. 
      
      Next, we compare the derived number counts in the different bands using our X-ray AGN population synthesis models.  
      The comparison between the {\it Chandra}  and the {\it Swift}/BAT number counts shows  a very good agreement for the 'standard' AGN spectrum with a power-law slope
      $\Gamma=1.9$, a high energy cut-off at $\sim$130 keV and a 2-10 keV reflection component of 3\%. 
      On the other hand, using the above standard AGN spectral model, only the bright part of the {\it NuSTAR} 8-24 keV and 3-8 keV number counts, 
      agree with the model predictions. 
      Then it is most likely that the disagreement between the X-ray number counts in the different bands 
      is because of the faint {\it NuSTAR} number counts. We discuss various possibilities for the origin of this disagreement.
    }
               
    \authorrunning{Akylas et al.}

    \maketitle
  %

	\section{Introduction}
	X-ray emission is a ubiquitous feature of Active Galactic Nuclei and X-ray surveys provide the most unbiased way for their detection.
	In contrast to the optical radiation, X-rays are not easily absorbed unless they encounter 
    extreme hydrogen column densities of the order of $\rm 10^{24}~cm^{-2}$. 
    Moreover, X-rays suffer from negligible contamination from stellar processes unlike the infrared and the optical emission. 
    Because of the above reasons, the X-ray surveys performed with {\it Chandra} and {\it XMM-Newton} in the 0.3-10 keV band, 
    have mapped very efficiently the AGN population and its evolution \citep{brandt2015}. These surveys have resolved more than 80\%  of the integrated X-ray light, the X-ray background, to Active Galactic Nuclei \citep{luo2017}. At harder energies $>$10 keV, where most of the X-ray background energy density is produced 
    \citep{revnivtsev2003, frontera2007, churazov2007, ajello2012}  our picture is far less clear.
    This  is mainly because of the lack of focusing X-ray instruments which could not allow for high sensitivity X-ray observations. 
    Nevertheless, the coded-mask instruments of the {\it INTEGRAL} and \swift missions have probed AGN at low redshifts, typically 
    $z<0.1$ \citep{malizia2009_int, ajello2012, vasudevan2013, oh2018}. 
  
    There is a drastic leap forward in our knowledge of the hard X-ray Universe 
    with the launch of the {\it NuSTAR} mission \citet{harrison2013}.
    This mission observes the X-ray sky  in the 3-80 keV band with 
    an unprecedented sensitivity carrying the first X-ray telescope that focuses X-rays with energies above 10 keV.
    {The large effective area and the excellent (52 arcsec half-power diameter) spatial resolution 
    of  {\it NuSTAR} allow the detection of faint X-ray sources up to  two orders of magnitude     
    fainter than the faintest \swift - BAT sources detected. }
    The majority of the serendipitous {\it NuSTAR} sources in the 8-24 keV band,  {are  associated
    with AGN at  redshifts of $ z=0.5-0.7 $ \citep{lansbury2017}}. 
     Although {\it NuSTAR} has an excellent sensitivity, its field 
     of view has a moderate size ($\sim$12 arcmin diameter). Then the \swift 
     and the {\it INTEGRAL} missions are quite complementary to {\it NuSTAR} as only these can observe large 
     swaths of the sky and hence large cosmological volumes. 
     
     \citet{harrison2015} presented the {\it NuSTAR} X-ray numbers counts in the 8-24 keV band. 
     These reach fluxes as faint as $2\times 10^{-14}$ \funits, resolving 33-39\% of the X-ray background 
     in this energy band. The measured {\it NuSTAR} counts lie significantly above a simple extrapolation with a Euclidean slope 
     of the {\it Swift}/BAT number counts measured at higher fluxes. \citet{harrison2015} assert that this may 
	suggest strong AGN  evolution between the average redshift of the BAT AGN ($z<0.1$) and the redshift of the \nustar sources 
	($z \sim 0.7$). \citet{aird2015a} derived the \nustar AGN luminosity function in the 8-24 keV band.    
	They find an excess number of sources relative  
	to  the BAT luminosity function. In contrast, the luminosity function of \citet{ueda2014} in the 2-10 keV band 
	agrees well with the BAT luminosity function \citep{ajello2012} assuming a spectrum with a reflection
	parameter of $R\sim1$ \citep[see][]{Zdziarski1995}  which corresponds to a 5\% fraction of reflected emission in the 2-10 keV band \citep[e.g.][]{akylas2012}. 
	\citet{aird2015a} suggest that a way to bring in agreement the \citet{ueda2014} and the \nustar luminosity function is to increase 
	the reflection parameter to $R\sim2$, corresponding to a reflection fraction of 10\% in the 2-10 keV band. 
	However, this leaves in disagreement the BAT
	with both the \nustar and the \chandra luminosity functions.

	Here, we address anew this problem by deriving the number count distributions from the newly released 105-month BAT survey (14-195 keV) and 
	the \nustar serendipitous source catalogue of \citet{lansbury2017} in both the 3-8 and 8-24 keV bands. We compare these distributions 
	using the X-ray population synthesis models of \citet{akylas2012}, assuming different AGN  spectral models, namely
	spectral indices, reflection component and high energy spectral cut-off. 
    
  \section{Data}   
 	
	\subsection{\swift}
	The {\it Swift} Gamma-Ray Burst (GRB) observatory \citep{gehrels2004} was launched in November 2004 and has been
	continually observing the hard X-ray ($\rm 14-195$ keV) sky with the Burst Alert Telescope (BAT). 
	BAT  is a large, coded-mask telescope, optimized to detect transient GRBs and is
	designed with a very wide field-of-view of $\sim$ 60$\times$100 degrees.

	The data presented in this paper stem from the analysis of the sources detected in the 105 months of observations 
	of the BAT hard X-ray detector on the \swift Gamma-Ray Burst observatory \citep{oh2018}. The 105 month BAT survey is a uniform, 
	hard X-ray, all-sky survey with a sensitivity of 8.40$\times10^{-12}$ ergs s$^{-1}$ cm$^{-2}$ over 90\% of the sky and 
	7.24$\times10^{-12}$ ergs s$^{-1}$ cm$^{-2}$ over 50\% of the sky, in the 14-195 keV band. The BAT 105 month 
	catalog provides 1632 hard X-ray sources in the 14-195 keV band above the 4.8$\sigma$ 
	significance level, with  422 new detections compared to the previous 70 month catalog release \citep{baumgartner2013}.	
	Our study is limited to the AGN population and therefore  all Galactic and extended sources have been excluded. 
	In particular, the following types of sources have been used: Seyfert I (379), Seyfert II (448), LINER (6), unknown AGN (114), 
	multiple (10), beamed AGN (158) and unidentified (129). Most of the 129 unidentified sources are located outside the 
	Galactic plane and therefore we expect that the majority of them are most likely AGN. 
  
  \subsection{NuSTAR}
	The Nuclear Spectroscopic Telescope Array, \nustar,  \citep{harrison2013} launched in June 2012, is the first
	orbiting X-ray observatory which focuses light at high energies (E $>$ 10 keV). It consists of two co-aligned focal
	plane modules (FPMs), which are identical in design. Each FPM covers
	the same 12 x 12 arcmin portion of the sky, and comprises of four Cadmium-Zinc-Tellurium detectors. 
	NuSTAR operates between 3 and 79 keV, and
	provides an improvement of at least two orders of magnitude  in sensitivity compared to
	previous hard X-ray observatories operating at energies E$>$10 keV.

	In our analysis, we  use the data from the first full catalogue for the \nustar serendipitous survey \citep{lansbury2017}.  
	The catalogue contains data taken during the first 40 months of the NuSTAR operation, with an area coverage of
	13 deg$^2$. For this study we use the 163 sources detected in the 8-24 keV energy range and the 273 sources detected in the 3-8 keV band. 
	We  remove all the sources that
	are associated with the primary science targets and the Galactic sources.
	Moreover, in order to further achieve the minimum contamination of the sample from non AGN sources, all detections within
       |b|$<$20 deg have been removed.  The final hard X-ray sample (8-24 keV) contains 106 sources and the soft X-ray sample (3-8 keV) 
	contains 171 sources. 
  

  \section{Number Count distribution}
  
  In this section, we derive the number count distributions, for the BAT (14-195 keV) and the \nustar (both the 8-24 and the 3-8 keV bands) data. 
  {The area curves have been taken from \citep{oh2018} and Lansbury (priv. comm.) for the BAT and the \nustar observations respectively. We use a   
  novel C-stat estimator which makes use of the un-binned data to fit the derived distributions.}
  {We  compare our findings in the hard 8-24 keV band with previous \nustar estimates and in the soft 3-8 keV \nustar band with precise measurements  of the \chandra number  count distribution in the 2-10 keV band.}
  Finally, we compare with the predictions of X-ray population synthesis models.

  \subsection{Methodology}
   
   In order to fit the number count distributions, we apply for the first time, the maximum likelihood statistic 
   for Poisson data, given in \citet{cash1979}. Normally, differential number count distributions are 
   binned in order to obtain at least 15 sources in each bin to apply Gaussian statistics. 
   However, the binning  may result in a loss of information.   
   \cite{cash1979} showed that the statistic estimator

  \begin{eqnarray}
  C=2\sum_{i=1}^{N} (e_i - n_i lne_i)
  \end{eqnarray}
  
  allows bins to be picked in an almost arbitrary way, while $\delta C=C-C_{min}$, {($C_{min}$ is the minimum value of the estimator)},   
  is distributed as $\delta \chi^2$ with  q  degrees of freedom and the same technique for generation of confidence intervals 
  could apply in the Poisson limit. In the equation above, N is the finite number of bins, $e_i$ is the expected number 
  (predicted  by the model)  and $n_i$ is the observed number in the $i^{th}$ bin. When one takes a  
  very fine mesh of bins, $n_i$ becomes zero or one and the statistic takes the form:

    \begin{eqnarray}
    C=2(E-\sum_{i=1}^{n} lne_i)
    \label{C}
    \end{eqnarray}
    
    where E is the total expected counts from the experiment and the summation is now over each of the observed photons.
   We assume that the differential LogN-LogS distribution is described  by a single power-law model, with slope $\beta$ i.e. $\rm log_{10} (dN/dS) = k + \beta log_{10} (S)$ and therefore we neglect the break of the number counts that normally appears around
   a 2-10 keV flux of $10^{-14}$ \funits \citep{age2008}. Hereafter, we quote the 
   slope of the integral number counts, $\alpha$, in order to facilitate the comparison with other results mentioned in the literature.
   The slope of the integral number counts $\alpha$   is related to the slope of the differential number counts $\beta$ with the relation  $\alpha=\beta-1$. 
  
   We calculate the minimum value of the C estimator (equation \ref{C}) using a very fine grid for the values of the slope $\beta$ and the 
  normalization parameter k, to obtain the best fit solution. Then we calculate the 90 per cent confidence intervals for two interesting parameters 
  by applying the criterion $\rm \delta C = C - C_{min} = 4.61 $ \citep{press2007}.   
  
  \begin{table*}
   \caption{Summary of the integral LogN-LogS results.} 
   \centering
   \begin{tabular}{cccc}

   \hline
   Mission & Energy band (keV) &  slope & Reference\\ 
   \hline
   \nustar & 3-8  & 1.82$\pm$ 0.15   & this work \\
           &      & 1.81$\pm$ 0.08   &  \cite{harrison2015} \\
   \nustar & 8-24 & 1.71$\pm$ 0.20   & this work \\
           &      & 1.76$\pm$ 0.10   & \cite{harrison2015} \\
           &      & 1.36$\pm$ 0.28   & \cite{zappa2018} \\
   \swift  & 14-195 & 1.51$\pm$ 0.10 & this work  \\
           &        & 1.42$\pm$ 0.14 & \cite{tueller2008}  \\
   \chandra & 2-10  & 1.52$\pm 0.07$ & \cite{age2008} \\
   \hline
   
   \end{tabular}
   \end{table*}

  \subsection{\swift 14-195 keV}

\begin{figure*}
    \begin{center}
    \includegraphics[height=0.6\columnwidth]{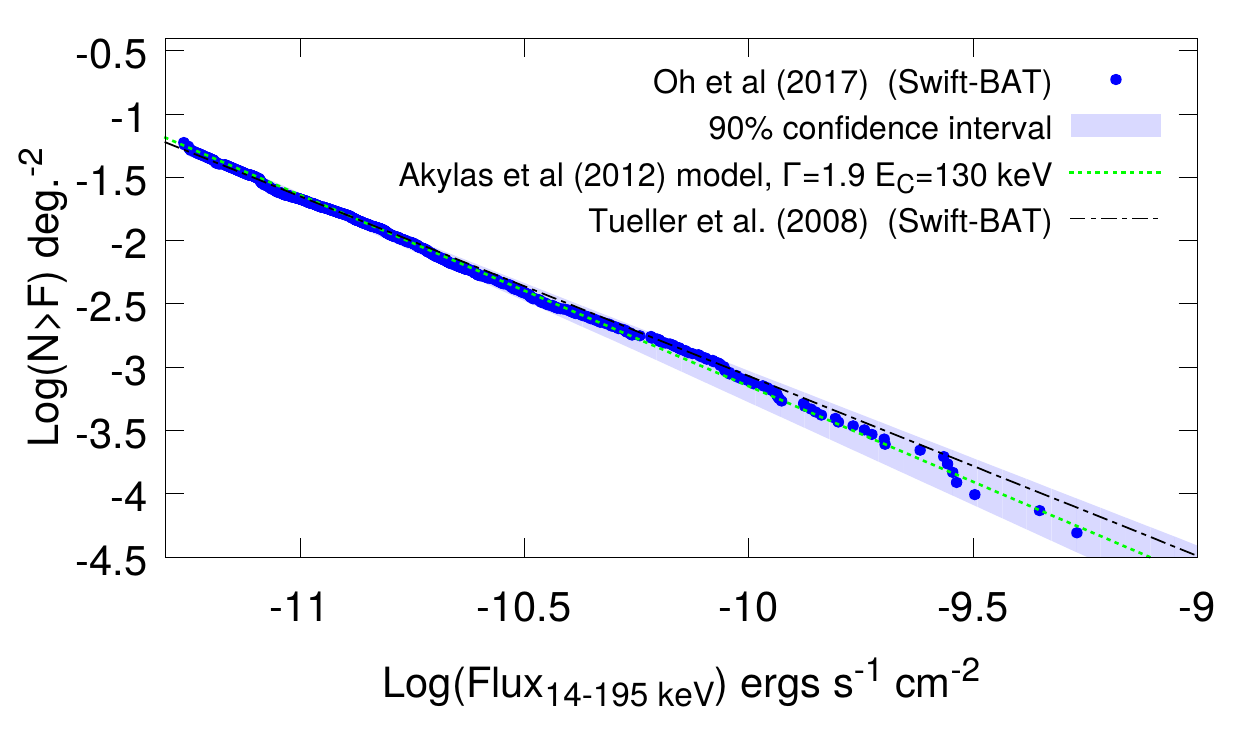}
        \includegraphics[height=0.6\columnwidth]{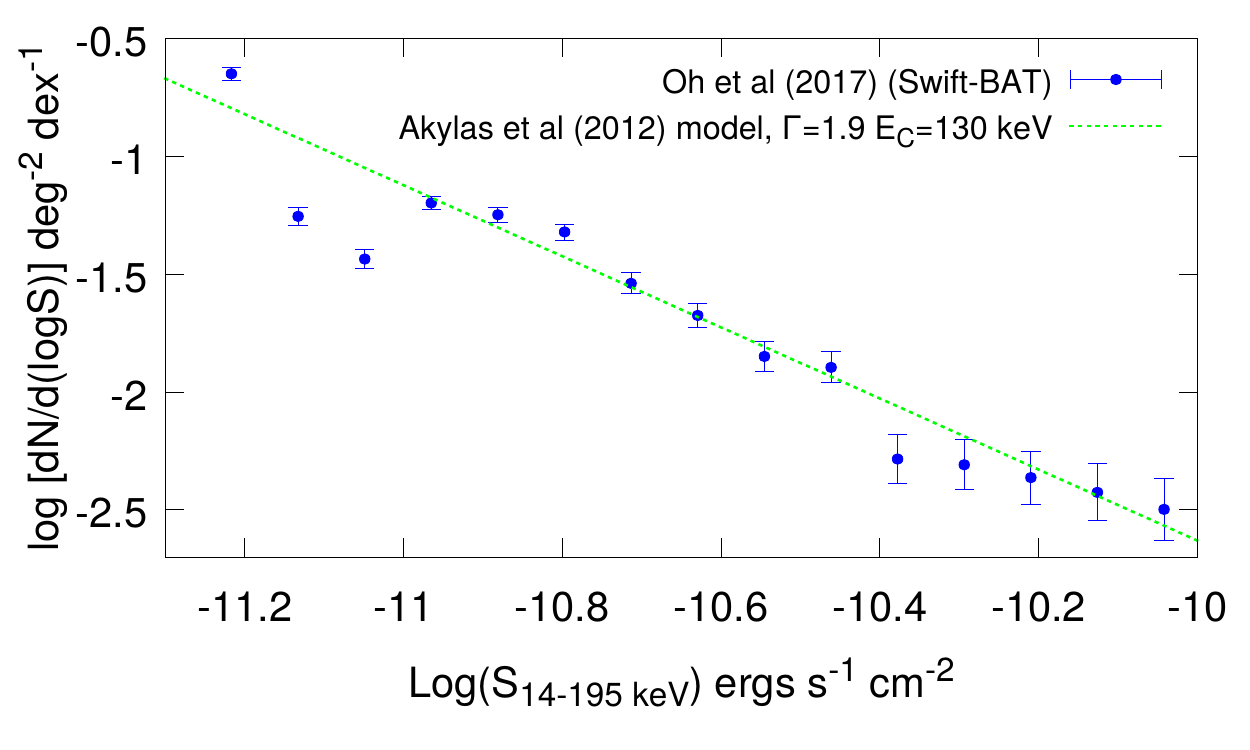}
    \end{center}
    \caption{{Right panel: The differential number count distribution in the 14-195 keV band for the BAT sample (blue points). The green dotted line is a characteristic model that fits the data at the 90\%.
    Left panel: The integrated number count distribution. The shaded area defines the 90\% confidence interval. The long-short dashed line shows previous results of \citet{tueller2008}.}  }
    \label{lognlogs14-195}
    \end{figure*}

  In Fig. \ref{lognlogs14-195} {we plot the differential and the integrated number count distributions} in the 14-195 keV band, for the BAT sample. 
  The best fit slope of the derived LogN-LogS and its 90\% uncertainty,
  is $\rm \alpha=1.51\pm 0.10$ {and the normalization $\rm log(k)=-18.22 \pm 0.63$}. 
  Similar findings  have been presented in  \cite{tueller2008}. 
  These authors compiled a sample of 103 AGN from the first 9-month  BAT survey. 
  They fit a power-law to their logN-logS and found a slope of 1.42$\pm$0.14.  
  
  Next, we use the X-ray population synthesis models presented in \citet{akylas2012},  to estimate the 
  average AGN spectral parameters that are consistent with the observed number count distribution. 
  We create a set of logN-logS predictions letting  the photon index, $\Gamma$, and the high energy cut-off to vary freely,
  while we fix  the amount of the reflected emission to 3\% of the total 2-10 keV luminosity.  
  The reflected emission has a small impact on the number count distribution of the total AGN population and the selected value of 3\% 
  is  typical of those reported in the literature (e.g. \cite{delmoro2017, ricci2017, zappa2018}). 
  A characteristic model  presented in Fig. \ref{lognlogs14-195} corresponds to $\Gamma=1.9$ and $\rm E_C$=130 keV  
  and is consistent with the data at the 90\% confidence level. 
  This value of the cut-off is entirely consistent with the {\it INTEGRAL} observations of 41 type-1 Seyfert-1. \citet{malizia2014}
  finds $E_C=128$ keV with a standard deviation of 46 keV. More recent work on the cut-off energy using \nustar data  \citep{tortosa2018}
  is in overall agreement with our work. 
  We note that when we adjust the reflection parameter to 1\% and 5\%, then a model consistent with the observed logN-logS, is obtained for $\Gamma=1.9$ and $\rm E_C$=200 keV or $\rm E_C$=80 keV respectively.

   \subsection{NuSTAR 8-24 keV}
 
\begin{figure*}
    \begin{center}
    \includegraphics[height=0.60\columnwidth]{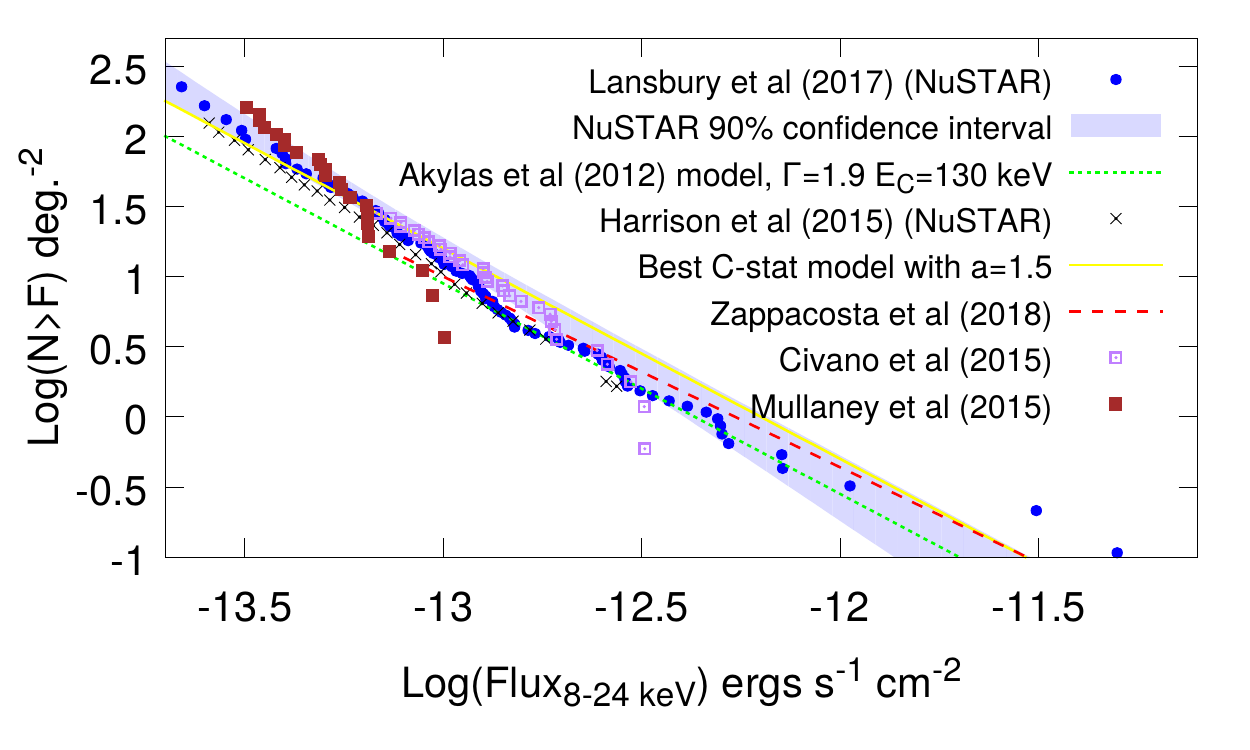}
    \includegraphics[height=0.60\columnwidth]{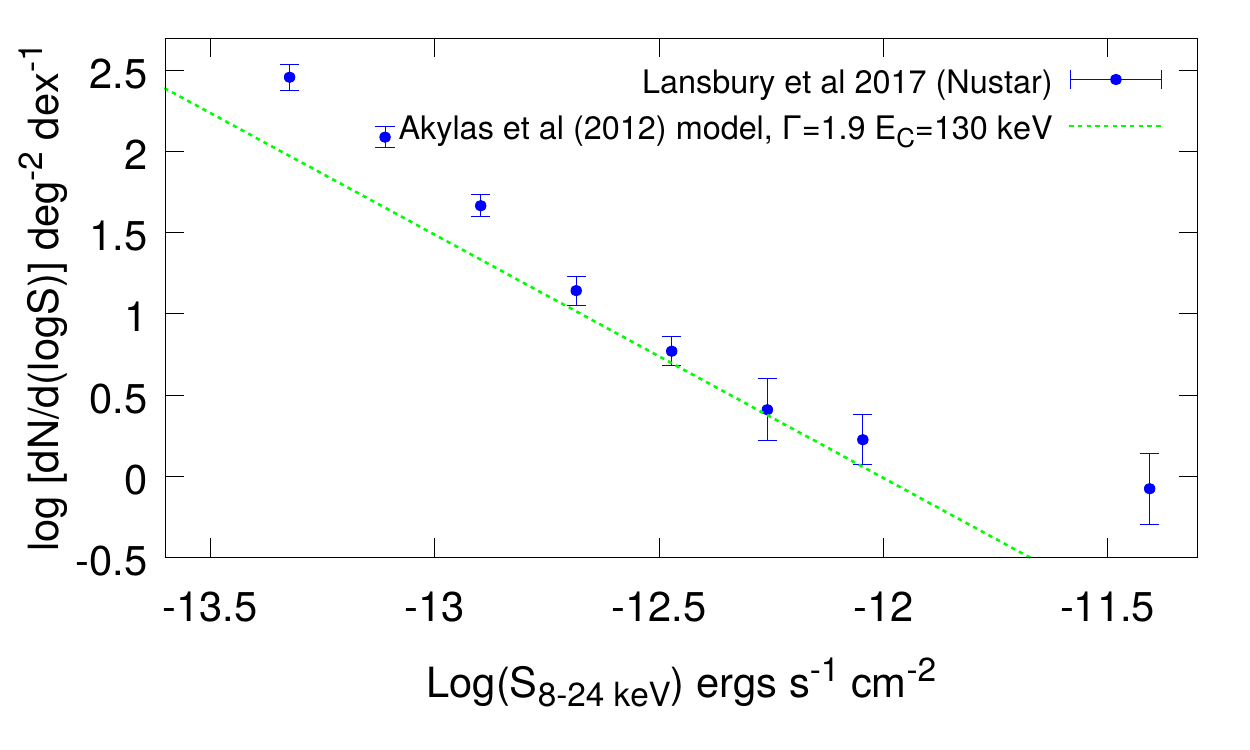}
    \end{center}
    \caption{{Right panel: The differential number count distribution in the 8-24 keV band for the \nustar sample. The green dotted line corresponds to the same characteristic model plotted in Fig. \ref{lognlogs14-195} computed now in the 8-24 keV band.  
    Left panel: The integrated number count distribution. The shaded area defines the 90\% confidence interval. The yellow solid line is the best C-stat solution to the \nustar counts, with a slope fixed to 1.5. The black crosses represent the \nustar results presented in \citet{harrison2015} and the long dashed red line shows the estimates of \citet{zappa2018}. {Also for comparison we show the results from \citet{mullaney2015} (red filled squares) and \citet{civano2015} (violet open squares).}}}
    \label{lognlogs8-24}
    \end{figure*}

    In Fig. \ref{lognlogs8-24}, {we plot the differential and the integrated number count distributions} of the sources detected  in the \nustar serendipitous 
	catalogue in the 8-24 keV band.  The best fit power-law slope  and its 90\% uncertainty  is $\rm \alpha=1.71\pm 0.20$ {and the normalization $\rm log(k)=-21.04\pm 1.51$}.
    This result is marginally consistent with the canonical Euclidean value of 1.5. 
    We compare our results with those presented in \citet{harrison2015}.  These authors measured the 8-24 keV number count 
	distribution by combining data from the following \nustar surveys:  (a) serendipitous source catalogue, 
    (b) COSMOS survey, \citet{civano2015}, (c) Extended Chandra Deep Field South, ECDFS, \citealt{mullaney2015} and (d) the 
    Extended Groth Strip region (EGS, Aird et al in prep.). Their best-fit  power-law, is very similar to our result with 
	$\rm \alpha=1.76\pm 0.10$. {Our estimates are 10 per cent higher in terms of the normalization, 
	at the flux of 10$^{-13}$ \funits. Note however that in this work we consider the full serendipitous survey catalog presented 
	in \citet{lansbury2017}, while \cite{harrison2015} incorporated a smaller part of the survey.} Their results are included in Fig. \ref{lognlogs8-24} for comparison. 
            
    In the same figure, we plot the prediction from the population synthesis model presented in Fig. \ref{lognlogs14-195} 
	for the BAT observations, calculated now in the 8-24 keV band. 
    There is a reasonable agreement between the model and the data in the bright part of the number count distribution 
    ($\rm f_{8-24keV}\sim2\times10^{-13}$ \funits) while 
    our model is way below the number count distribution at fainter fluxes.
    Recently, \cite{zappa2018}, analyzed a bright sample of 63 sources ($\rm f_{8-24 keV}>7\times10^{-14}$ \funits) from the 
	multi-tiered \nustar Extragalactic Survey program. 
    Their estimates on the number count distribution suggest a flatter slope ($\rm \alpha=1.36\pm 0.28$) but still consistent 
	with the Euclidean slope of $\alpha$=1.5 within the errors.
	In Fig. \ref{lognlogs8-24} the red dashed line shows their results. 
    There is a notable agreement between their estimates and our results.    
    Finally, we give the best C-stat solution to the \nustar counts, with a slope fixed to 1.5, to enable the direct comparison 
	of our predictions with the observed counts. 
	In this case the normalisation of the logN-logS becomes about a factor of two higher than our model.

    \subsection{NuSTAR 3-8 keV}
    
\begin{figure*}
   \begin{center}
   \includegraphics[height=0.6\columnwidth]{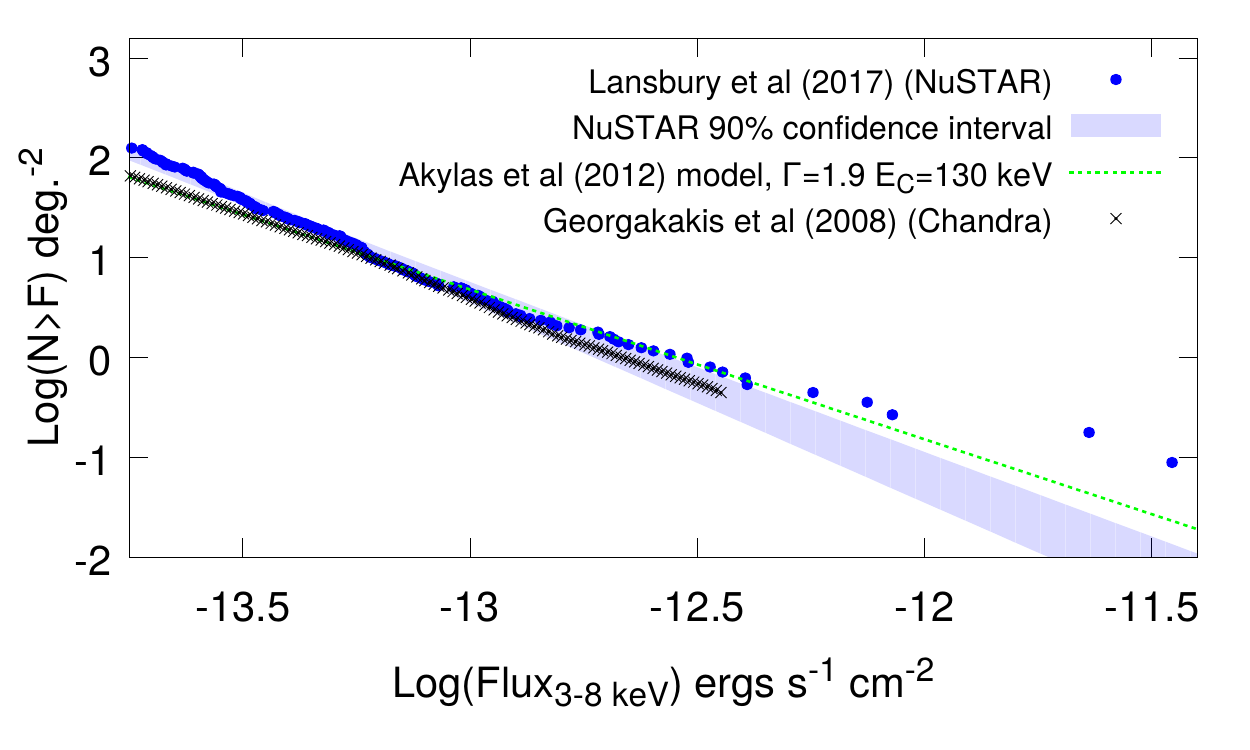}
   \includegraphics[height=0.6\columnwidth]{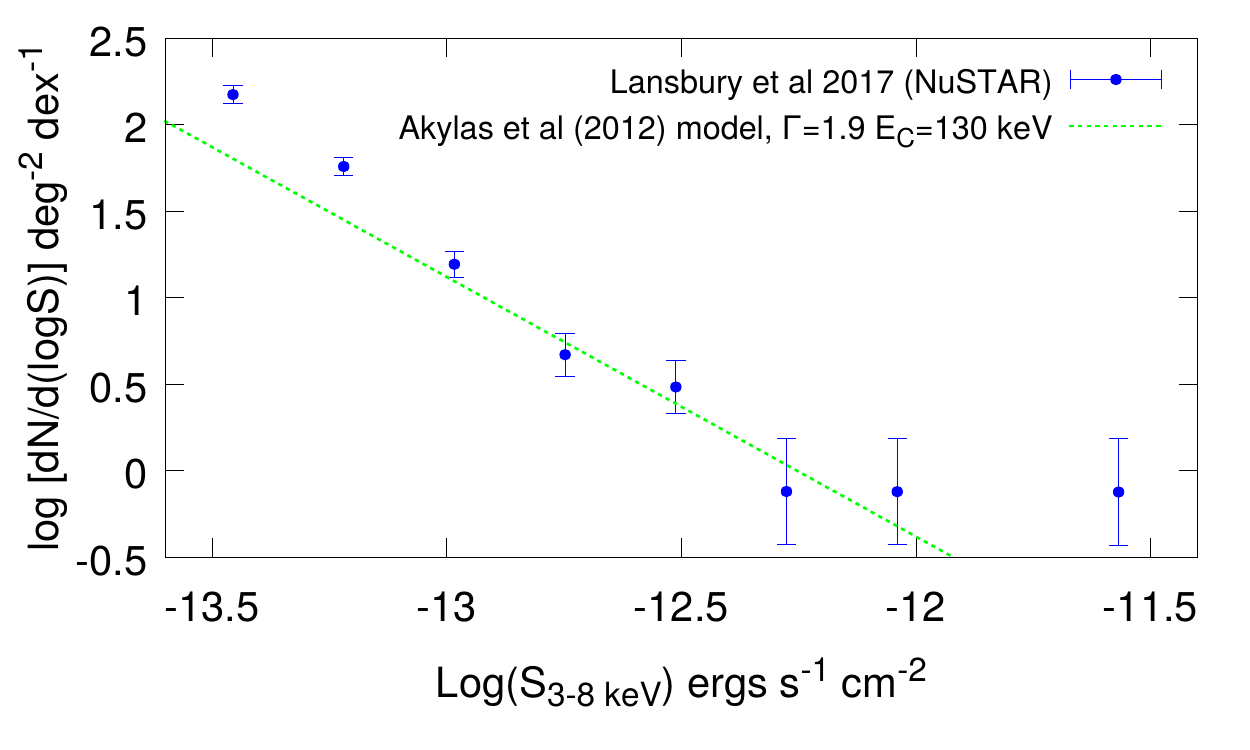}
   \end{center}
   \caption{{Right panel: The differential number count distribution in the 3-8 keV band for the \nustar sample. The green dotted line corresponds to the characteristic model plotted in Fig. \ref{lognlogs14-195}, computed here in the 3-8 keV band.
   Left panel: The shaded area defines the 90\% confidence interval based on the C-stat minimization estimator. The crossed points show the number count distribution of \citet{age2008}.}}  
   \label{lognlogs3-8}
   \end{figure*}

    In Fig. \ref{lognlogs3-8} {we plot the differential and the integrated number count distributions} in the 3-8 keV band using the \nustar serendipitous source catalogue. 
    The best fit slope of the distribution and its 90\% uncertainty, is $\rm \alpha=1.82\pm 0.15$ {and the normalization $\rm log(k)=-22.98\pm1.22$}. 
    The observed distribution is consistent with the canonical Euclidean value of 1.5, at the 99.5\% confidence interval. 
    This estimate is also in excellent agreement with the slope  found by \citet{harrison2015}, 
    $\rm \alpha=1.81\pm 0.08$. 
    
    In the same figure we  plot the model presented in Figs. \ref{lognlogs14-195} and Fig. \ref{lognlogs8-24} 
    (green dotted line). Similarly to the \nustar results in the 8-24 keV band, there is a marked difference between our models 
    and the \nustar data at faint fluxes ($<7\times10^{-14}$ \funits) while 
    there is good agreement at brighter fluxes.    
    
    Next, we include in our comparison the 2-10 keV \chandra number counts presented in \citet{age2008}. 
    These authors, used a novel technique which correctly accounts for the observational biases that affect the probability of 
    detecting a source of a given X-ray flux.
    They estimated the X-ray source counts by combining deep pencil-beam and shallow wide-area \chandra observations. 
    Their sample has a total of 6295 unique sources over an area of 11.8$\rm deg^{2}$. 
    The flux conversion from their 2-10 keV band to the 3-8 keV  adopted here, has been made 
    assuming a simple power-law model with $\Gamma=1.9$. Because of  the limited energy width of the bandpasses, 
    the choice of the spectral slope has  a small impact on the conversion from the 2-10 keV to the 3-8 keV band.
    Similarly to the 8-24 keV results, the \nustar number count distribution nicely matches the \chandra data only at the brighter fluxes.
    It is also evident from the plot that the \chandra  number counts are consistent with the same standard model that also describes very well the BAT logN-logS.

   \section{Summary}
 
   The hard X-ray sky above 10 keV has been observed in unprecedented sensitivity with the \nustar mission. 
   A difference between the BAT (14-195 keV) and the 8-24 keV \nustar counts has been reported in the literature in the sense that 
   the former lies well below the latter. In the light of this disagreement, 
   we estimate the number count distribution from the recently released 105-month {\it Swift}-BAT AGN catalogue, in the 14-195 keV band. 
   For comparison we also derive the \nustar number counts from the serendipitous source catalogue of \citet{lansbury2017}.
   We use a novel approach for fitting the number count distribution, by applying the C-statistic estimator on the un-binned data.
   We compare the number count  distributions between  different bands  using our X-ray AGN population synthesis models. 
   The results are presented in Table 1, and can be summarized as follows:
   
   1. The BAT logN-logS presents a Euclidean slope of $\alpha=1.51\pm0.10$.   This result is consistent  with earlier estimates of the
   BAT logN-logS \citep{tueller2008} which used considerably fewer sources. 
   
   We used our X-ray AGN population synthesis  code \citep{akylas2012}, to find  physically  motivated models that  fit these observations. 
   A 'standard' AGN spectral model with a photon index 1.9, a high energy cut-off 130 keV, and reflected emission of 3$\%$ of the total 2-10 keV 
   flux matches very well the BAT number count distribution. 
      
   We also compared the BAT  number counts with the {\it Chandra} results in the softer, 2-10 keV band, obtained by \citet{age2008} 
   by estimating the above 'standard' model in the 2-10 keV band. Again, there is a notable agreement between our population synthesis model 
   predictions and the {\it Chandra} results. This demonstrates that reasonable assumptions on the spectral shape of AGN should  lead 
   to consistent modelling of the number count distributions in the {\it Chandra} and the BAT bands.  
   
   2. The \nustar number count distribution in the 8-24 keV band presents a slope of $\rm \alpha=1.71\pm 0.20$. This is steeper than the Euclidean 
   slope but still consistent at the 90\% confidence level. Therefore, in order to facilitate the comparison, we fix the slope to 1.5. 
   Then it is evident that the {\it NuSTAR} number count distribution lies above the  BAT counts by about a  factor of two. It is evident 
   from Fig. \ref{lognlogs8-24}  that the same population synthesis model, that successfully  describes both the BAT and the 
   {\it Chandra} observations, agrees  only with the bright end of the {\it NuSTAR} logN-logS. This result is further confirmed by recent findings 
   of \citet{zappa2018}. Their  estimates on the number count distribution of a bright sample ($>7\times10^{-14}$ \funits) from the NuSTAR 
   Extragalactic Survey program, nicely agree with both our logN-logS and the predictions of our 'standard' model in the 8-24 keV band. 
   
   3. The \nustar number count distribution in the soft 3-8 keV band presents a slope $\rm \alpha=1.82\pm 0.15$, 
    significantly steeper than the Euclidean slope at the 99.5\% confidence level.   
    Similarly to the results in the 8-24 keV band, only the brightest part of the \nustar soft  number count distribution is compatible 
	with the \chandra results and the predictions of our 'standard' model. 
	
	Instead of using the number count distributions, 
	\citet{aird2015a} compared the luminosity function derived in the following bands: \chandra 2-10 keV, \nustar 8-24 keV and BAT 14-195 keV. 
	Assuming a reflection fraction of 5\% in the 2-10 keV band they find a good agreement between the BAT and the \chandra luminosity function, 
	while \nustar is way above the previous two luminosity functions. When they assume a much stronger reflection component of 10\%, 
	the \nustar 8-24 keV and the \chandra luminosity functions  come in agreement. However, this comes at the cost of the BAT luminosity function which remains
	way below the other two. Using our number count modelling instead, we suggest that this disagreement could be circumvented only by assuming a  
	rather extreme AGN spectral model, with a power-law high energy cut-off of less than 50 keV and at the same time a strong 
	reflection component (reflection fraction greater that 10\% in the 2-10 keV band).
	This is because the increased reflection component significantly enhances the flux in both the 8-24 keV and the 14-195 keV 
	bands, while the low energy cut-off cancels this gain only in the 14-195 keV band. 
    Thus the net result is a considerable increase in the 8-24 keV band. However, even in this case, the major problem remains 
    in the comparison  between the \nustar 3-8 keV and the \chandra 2-10 keV number count distributions. 
    
   {Interestingly, this does not appear to be only a problem of the serendipitous survey. 
    The logN-logS in both the COSMOS and the CDFS fields also present an abrupt upturn albeit at different fluxes. 
    The above findings may suggest that  a 
    fraction of spurious sources has been included at faint fluxes. 
     However, this is not very  likely given the high fraction $\sim85\%$ of soft X-ray   detected counterparts of the \nustar sources,  
    which does not vary significantly with flux. 
    Alternatively the \nustar area curve could be  underestimated at faint fluxes i.e. near the centre of the field-of-view where the sensitivity is the highest. 
    As can bee seen from  Fig. 2, at the faintest flux bin the area curve in the 8-24 keV band should have been underestimated by a factor of about 2.5 .    
    Finally, the possibility that the Eddington bias plays some role cannot be ruled out. It is unclear however why the Eddington bias 
    affects fluxes a factor of four brighter than the survey's flux limit, given that the flux errors are moderate i.e. of the order of 10\%.}

  \begin{acknowledgements}
  We would like to thank the anonymous referee for many suggestions that helped to improve the 
  paper. We are also grateful to George Lansbury and James Aird for many useful discussions. 
  \end{acknowledgements}

\bibliography{ref}{}
\bibliographystyle{aa}
  \end{document}